\documentclass[structabstract]{aa}
\usepackage{graphicx}
\usepackage{txfonts}
\usepackage{natbib}        
\bibpunct{(}{)}{;}{a}{}{,} 

\def\Ha{\mbox{H$\alpha$}}
\def\Rstar{$R_{\star}$}

\def\Rsun{\mbox{${R}_{\odot}$}}

\def\kms{$\mathrm{km\,s^{-1}}$}

\begin{document}
\title{New activity in the large circumstellar disk \\ 
                          of the Be-shell star 48\,Lib
    \thanks{Based partly on observations collected at the
    European Southern Observatory, Chile (Prop.\ Nos. 383.D-0522, 266.D-5655, 
    081.C-0475).}
       }
   \author{S.~\v{S}tefl \inst{1}
             \and 
            J.-B.~Le~Bouquin \inst{2} 
             \and  
            A.~C.~Carciofi \inst{3}
             \and 
             T.~Rivinius \inst{1}
             \and
            D.~Baade \inst{4}
              \and  
            F.~Rantakyr\"o \inst{5} 
}
\offprints{S.\ \v{S}tefl, \email{sstefl@eso.org}}
\institute{European Organisation for Astronomical Research in the Southern
           Hemisphere, Casilla 19001, Santiago 19, Chile 
           \and
           UJF-Grenoble 1 / CNRS-INSU, Institut de Plan\'etologie et d’Astrophysique
           de Grenoble (IPAG) UMR 5274, Grenoble, France
           \and
           Instituto de Astronomia, Geof{\'\i}sica e Ci{\^e}ncias
           Atmosf{\'e}ricas, Universidade de S\~ao Paulo, Rua do Mat\~ao 1226, 
           Cidade Universit\'aria, S\~ao Paulo, SP 05508-900, Brazil 
           \and 
           European Organisation for Astronomical Research in the Southern 
           Hemisphere, Karl-Schwarzschild-Str.~2, 85748 Garching bei M\"unchen, 
           Germany 
           \and
           Gemini Observatory,  Southern Operations Center, c/o AURA, Casilla 603, 
           La Serena, Chile
}
\date{Received: $<$date$>$; accepted: $<$date$>$; \LaTeX ed: \today}  
\authorrunning{\v{S}tefl et al.}
\titlerunning{New circumstellar-disk activity in 48\,Lib}
\abstract
{}
{Spectroscopic, polarimetric, and high spectral resolution interferometric 
data covering the period 1995--2011 are analyzed to document the transition 
into a new phase of circumstellar disk activity in the classical Be-shell star 
48\,Lib.  The objective is to use this broad data set to additionally test 
disk oscillations as the basic underlying dynamical process.  
} 
{The long-term disk evolution is described using the $V/R$ ratio of the violet 
and red emission components of H$\alpha$ and Br$\gamma$, radial velocities and 
profiles of \ion{He}{i} and optical metal shell lines, as well as multi-band 
BVRI polarimetry.  Single-epoch broad-band and high-resolution interferometric 
visibilities and phases are discussed with respect to a classical disk model 
and the given baseline orientations.}
{Spectroscopic signatures of disk asymmetries in 48\,Lib vanished in the late 
nineties but recovered some time between 2004 and 2007, as shown by a new 
large-amplitude and long-duration $V/R$ cycle.  Variations in the radial 
velocity and line profile of conventional shell lines correlate with the 
$V/R$ behavior.
  They are shared by narrow absorption cores superimposed on otherwise 
seemingly photospheric \ion{He}{i} lines, which may form in high-density gas 
at the inner disk close to the photosphere.  Large radial velocity 
variations continued also during the $V/R$-quiet years, suggesting that 
$V/R$ may not always be a good indicator of global density waves in the disk.  
The comparison of the polarization after the recovery of the $V/R$ activity 
shows a slight increase, while the polarization angle has been constant for 
more than 20 years, placing tight limits on any 3-D precession or warping 
of the disk. The broad H-band interferometry gives a disk diameter of 
(1.72$\pm$0.2)\,mas  (equivalent to 15 stellar radii),  
position angle of the disk (50$\pm$9)\degr  and a relatively low disk 
flattening of 1.66$\pm$0.3. Within the errors the same disk position angle is  
derived from polarimetric observations and from photocenter shifts across 
Br$\gamma$. The high-resolution interferometric visibility and phase profiles 
show a double or even multiple-component structure. A preliminary 
estimate based on the size of the Br$\gamma$ emitting region indicates a 
large diameter for the disk (tens of stellar radii).
Overall, no serious contradiction between the observations and 
the disk-oscillation model could be construed.}
{}
\maketitle
%

\section{Introduction}
\label{intro}

Circumstellar matter around rapidly rotating classical Be stars has been studied 
since the discovery by \citet{1866AN.....68...63S} of emission lines in 
$\gamma$\,Cas.  
Only some 60 years later, \citet{1931ApJ....73...94S} laid the foundation 
to the notion of a flattened structure of this matter.  It took again more 
than 60 years until the disk nature was unambiguously proved by the combination 
of polarimetry and interferometry  \citep[e.g.][]{1997ApJ...479..477Q}. 
For a general overview of classical Be stars and their properties see 
\citet{2003PASP..115.1153P}. 

During the 
last decade, the rate of progress was much accelerated by the availability for 
detailed modeling of combined high-resolution spatial and spectral observations.  
This concerns especially the disk dynamics.  Perhaps the most manifest empirical 
indicator of dynamical processes in Be star disks is the so-called $V/R$ ratio 
of the strengths of violet and red emission-line components.  The nature of this 
variability has long remained somewhat enigmatic.  Today, it is widely accepted 
that the long-lived and frequently quasi-periodic structural variations of emission 
lines find their basic explanation in one-armed density waves of the disk 
\citep{1997A&A...318..548O,2009A&A...504..915C}.  Although the non-periodicity of 
the variability makes it difficult to distinguish systematic residuals from 
ephemeral events, the study of density waves and other phenomena triggered by 
the presence of a stellar companion has a strong potential 
for revealing the structure, dynamics,  and secular evolution of Be disks 
\citep{2011IAUS..272..325C}. 

The activity and even the existence of Be disks typically vary on timescales of 
years, including their temporary absence.  The gas of the disk is both provided 
and dissipated by the central star, with the disk build-up often resulting from 
discrete stellar mass-loss events/episodes 
\citep[e.g.,][]{2011IAUS..272....1B,2011arXiv1112.0053C}. It is assumed that the 
rotationally induced quadrupole moment of the stellar gravitational potential is 
the driving force of the disk oscillations \citep{1992A&A...265L..45P}.  In view 
of extended periods without observable $V/R$ activity, this invites the question 
whether variable star-to-disk mass transfer or restructurings of the disk 
\citep[e.g., to a more ring-like profile,][] {2001A&A...379..257R} play a role 
in triggering or quenching density waves.   

\begin{table*}[t]
\begin{minipage}{175mm}
\begin{center}
\caption[]{New and archival 48\,Lib spectroscopic datasets in the visual 
           and near-IR range}
\begin{tabular}{cc@{\,--\,}cc@{\,--\,}cllrrc}
\hline\noalign{\smallskip}
\hline
\rule{0ex}{2.5ex} Data       & \multicolumn{2}{c}{Date} & \multicolumn{2}{c}{JD}       &                                     &                                    & 
 \multicolumn{1}{c}{No.\ of} & \multicolumn{1}{c}{Resolving} & \multicolumn{1}{c}{Spectral}   \\
 
set                          & \multicolumn{2}{c}{}     & \multicolumn{2}{c}{2400000+} & \raisebox{1.5ex}[1.5ex]{Telescope}  & \raisebox{1.5ex}[1.5ex]{Instrument} & 
\multicolumn{1}{c}{spectra}  &\multicolumn{1}{c}{power} & \multicolumn{1}{c}{range [\AA]}  \\ 
\hline \noalign{\smallskip}
A & 1995        &       2003         & 49\,788   & 52\,725     & ESO 50-cm, Ond\v{r}ejov 2-m   & {\sc Heros} &  18+18  & 20000 & 3500--5540, 5800--8600  \\ 
B & \multicolumn{2}{c}{2001-07-11} & \multicolumn{2}{c}{52\,101} & ESO/VLT-Kueyen & {\sc UVES} &   4  & 60\,000 & 3100--10400  \\
C & \multicolumn{2}{c}{2008-04-03} & \multicolumn{2}{c}{54559} & ESO/VLT-Kueyen  & {\sc UVES} &   9  & 60\,000 & 4800--6800  \\
D & 2007 & 2011              & 54\,172   & 55\,389                & BeSS database   & {\sc LHIRES} &  19  & 17\,000 & 6520--6610  \\ 
E & \multicolumn{2}{c}{2009, 2011} & \multicolumn{2}{c}{55\,014,55\,695} & OPD/LNA         & {\sc ECASS} & 4+12  & 16\,000 & 3950--4900, 6070--6855  \\ 
\hline \noalign{\smallskip}
F & \multicolumn{2}{c}{2009-08-06} & \multicolumn{2}{c}{55\,049.5445} & Gemini South & {\sc Phoenix} &  1  & 45\,000 & 21\,600--21\,700  \\
\hline \hline  \noalign{\smallskip}
\end{tabular}
\end{center}
\end{minipage}
\label{datasets}
\end{table*}

\begin{table*}
\caption[]{High-resolution {\sc AMBER} and broad-band {\sc PIONIER} observations of 48\,Lib 
 and $\zeta$\,Tau. The elementary integration time DIT is given in seconds per frame for {\sc AMBER} 
but by scanning frequency in fringe per second for {\sc PIONIER}.
}
\begin{tabular}{llrrrrrr}
  \hline\noalign{\smallskip}
  \hline
  Date       &      JD          &    Target   & Instrument   &  Baseline     & DIT         &  Spectral  &  Spectral   \\
             & (+ 2\,400\,000)  &             &              & configuration &             &    band    &  resolution \\
  \hline \noalign{\smallskip}
2011-06-04 &  55\,717.1998   & 48\,Lib      & {\sc PIONIER}& D0-G0-G1-I1  & $\sim$300\,Hz &   H       &   15        \\
2009-05-13 &   54\,963.2349   & 48\,Lib      & {\sc AMBER}  & UT1-UT2-UT4  &  3\,s        & Br$\gamma$ &  12\,000    \\
2008-11-16 &   54\,754.3453   & $\zeta$\,Tau & {\sc AMBER}  & UT1-UT2-UT4  &  1\,s        & Br$\gamma$ & 12\,000     \\
\hline\noalign{\smallskip}
\end{tabular}
\label{amber_data}
\end{table*}

Spectro-interferometry has a distinct potential to permit new insights to be 
gathered.
After the angular resolution of circumstellar disks by broad-band interferometry 
in the 1990s \citep{1997ApJ...479..477Q}, the increasing spectral resolution of 
interferometric equipment afforded detailed investigations also of the disk 
dynamics.  \citet{2007ApJ...654..527G} determined the size, inclination, and 
orientation of four northern Be-star disks by means of broad-band K observations 
with the CHARA array. Later, \citet{2007A&A...464...73M} detected an asymmetry in 
the disk of $\kappa$~CMa using the spectro-interferometric capabilities of 
{\sc AMBER} at medium resolution (MR, $R=1\,200$). MR {\sc AMBER} observations 
were also used for a multi-technique study of another bright Be star,  
$\zeta$\,Tau \citep{2009A&A...504..929S,2009A&A...504..915C}. 

In the second half of the 20th century, the H$\alpha$ $V/R$ ratio of 48\,Lib 
(HD\,142983, B3Ve, V~$\sim$~5.0) went through a number of irregular long-term 
cycles \citep[see, e.g.,][]{1995A&A...300..163H}.  Strong, narrow shell absorption 
lines show that the line of sight intersects the disk.  The large stellar 
$v \sin i$ of $\sim$400\,\kms \ implies that the disk is equatorial.  

\citet{2010ApJ...721..802P} used the spectro-astrometric mode of the Keck 
interferometer to study the structure of the 48\,Lib circumstellar disk. The 
derived disk size was 1.65 mas in the K-band continuum, estimated at about seven 
times the stellar diameter.  The disk was also resolved in several Pfund lines.  
Although 
the spectral resolution did not exceed 1\,000, the self-phase referencing mode 
permitted the authors to obtain differential astrometric precision of 3 
micro-arcsec.  
Pott et al. concluded that the continuum and the Pfund emission lines originate 
from a more compact region within the Br\,$\gamma$ emission zone, consistent with 
the fact that the Pfund lines are weaker than the Br\,$\gamma$ line.  They also 
inferred that deviations from axisymmetry are radius-dependent, which is 
characteristic of one-armed disk oscillations.  

One purpose of this report is to put the recently resumed emission line variability 
in the circumstellar disk of 48\,Lib into a common perspective with the general 
spectroscopic and polarimetric variability in the interval 1995--2011.  We also 
present results of the first {\sc AMBER} high-spectral resolution (HR, $R~=~12\,000$) 
interferometry of this star with a fairly complex appearance of interferometric 
visibilities and phases.  Similar HR observations of $\zeta$\,Tau \citep[not used 
by][]{2009A&A...504..929S} are included to demonstrate that this complexity 
may well not be a singular peculiarity but rather some commonality only revealed at 
higher spectral resolution.  The two stars also have significant long-term $V/R$ 
variations in common. 

Basic geometrical parameters of the disk are derived by fitting an axisymmetric 
Gaussian disk model to the {\sc PIONIER} H-band observations, but an exact 
dynamical modeling using {\sc AMBER} HR data and the HDUST code is postponed to the 
dedicated paper in preparation. A toy model that is unable to describe the 
dynamics of disk density waves would only be an unnecessary or even confusing 
half-step and is not included in this paper.

\section{Observations}
\label{data}

\subsection{Optical and near-infrared spectroscopy} 
\label{dataspectr}

To identify long-term variations and trends in the spectroscopic properties, 
data obtained with the {\sc Heros} spectrograph \citep{1998RvMA...11..177K} 
between 1995 and 2003 that are included in the statistical 
study by \citet{2006A&A...459..137R} were re-analyzed.  This echelle spectrograph 
covers the spectral region of 3500--8600\,\AA \ in 2 arms with a gap of about 
200\,\AA \ between them. During the observations of 48\,Lib, the spectrograph was 
attached to the ESO (La Silla) 50-cm and Ond\v{r}ejov 2-m telescopes. 
  
The ESO Science Archive only contains two sets of {\sc UVES} \citep{2000SPIE.4008..534D} 
echelle spectra, which were obtained on July 11, 2001 and April 3, 2008.  Although 
each of them is limited to one night, they enable a detailed comparison of spectral 
lines before and after the density wave became detectable again (see Fig.\ \ref{vtor}).  
Additional red and blue optical spectra at $R\,=\,16\,000$ were secured at the Pico 
dos Dias Observatory (OPD) of the Laborat\'orio Nacional de Astrof\'{i}sica (LNA), 
Brazil, on July 2, 2009 and May 12, 2011.  The observations were made with a 
Cassegrain spectrograph equipped with a 1200 grooves/mm grating blazed at 
6562\,\AA \ and a  1024x1024 pixel CCD.  To identify long-term \Ha \ 
$V/R$ trends in 2007--2011, we also used spectra from the BESS database 
(http://basebe.obspm.fr), which mainly comprises observations by amateur 
astronomers.  The spectral resolution of some BESS datasets is quite low but 
they define the $V/R$ variations well. 

In August 2009, a high-resolution ($R=45\,000$) spectrum of the Br$\gamma$ profile 
was obtained with the GEMINI/{\sc PHOENIX} echelle near-IR spectrograph 
\citep{2003SPIE.4834..353H}.  The spectrum covers the interval 2.16--2.17\,$\mu$m, 
and the 40\,s integration gave a signal-to-noise ratio $S/N\sim170$.

The characteristics of the spectra and their origin are summarized in 
Table~\ref{datasets}.

\subsection{Imaging polarimetry}
\label{datapolar}

New multi-band linear polarimetry was obtained in the period May 2009--September 
2010 using the IAGPOL imaging polarimeter attached to the Cassegrain focus of the 
0.6-m 
Boller \& Chivens telescope at OPD/LNA.  We used the CCD camera with a polarimetric 
module described in \citet{1996ASPC...97..118M}, employing a rotating half-wave 
plate and a calcite prism placed in the telescope beam.  A typical observation 
consists of eight consecutive wave-plate positions separated by 22\fdg5.  Details of 
the data reduction can be found in \citet{1984PASP...96..383M}.  In each observing 
run at least one of HD\,183143, HD\,145502, and HD\,187929 was observed as the
polarized standard star to calibrate the position angle.  

For comparison with the polarimetric status before the new $V/R$ cycle  
commenced  we used seasonal means in the U,B,V,R,I bands in 1990--1998 published by  
\citet{2001PhDT.......100M,1994PASP..106..949M,1999PASP..111..494M}. The values 
are based on observations obtained with the polarimeter described by 
\citet{1979ApJ...233...97B}, which is mounted on the University of Texas 0.9-m telescope 
at McDonald Observatory. For the period 1989--1994, they were complemented by 
V-band Halfwave Polarimeter data  \citep[HPOL;][]{1996ASPC...97..100N} from the 
0.9-m telescope of Pine Bluff Observatory, which were retrieved from the web site 
http://www.sal.wisc.edu/HPOL.

All polarimetric observations were corrected for the interstellar polarization of 
(0.819$\pm$0.050)\% and $PA=(96\pm5)^{\circ}$ derived by Draper \& Wisniewski 
(private communication) using HPOL spectropolarimetry and a Serkowsky law with 
$\lambda_\mathrm{max}=5591$\,\AA.

\begin{center}
   \begin{figure}[t]
   \includegraphics[width=\columnwidth]{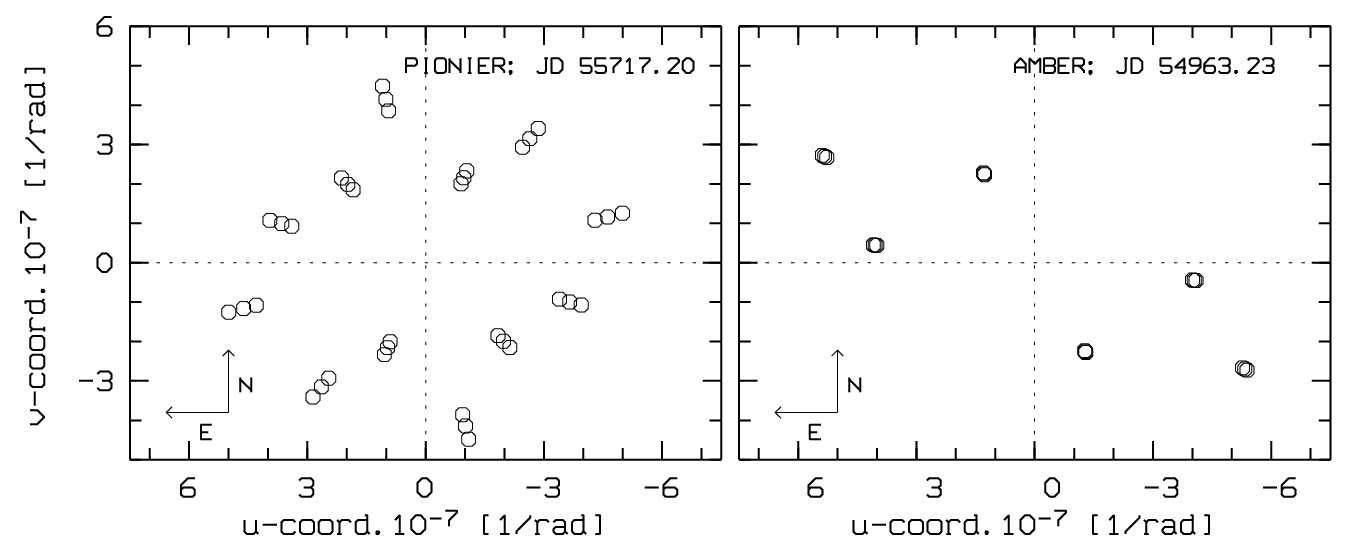}   
   \caption{u-v coverage of the 48\,Lib {\sc PIONIER} and {\sc AMBER} 
    observations. Spatial frequencies corresponding to the minimum, central, 
    and maximum wavelength are plotted for each baseline.
   }
\label{uvplots}
\end{figure}
\end{center}

\subsection{Interferometry}
\label{datainterfer}

Broad-band interferometric observations in the H band were obtained with the 
{\sc PIONIER} \citep{2011A&A...535A..67L} visitor instrument and the four 1.8-m auxiliary 
telescopes of ESO's VLTI system \citep{2003SPIE.4838...89G} on June 04, 2011. 
The AT configuration was D0-G1-H0-I1, with baselines ranging from 40\,m to 65\,m. The 
fringes were slightly dispersed over three spectral channels across the H band. The 
observing conditions were poor (DIMM \citep{1990A&A...227..294S} seeing 
$\sim$1.5--2.0\arcsec 
and coherence time $<$1\,ms). But we were able to reduce the data with the standard 
\texttt{pndrs} package described by \citet{2011A&A...535A..67L}.
The observations provide the simultaneous measurement of six absolutely calibrated 
visibilities and four closure-phases in the H-band.

High-spectral resolution (HR; $R=12\,000$) interferometric observations in the K 
band were obtained with the {\sc AMBER} instrument \citep{2007A&A...464....1P} and 
the 8-m UT telescopes of the VLTI on May 12, 2009.  
For comparison, {\sc AMBER} observations of $\zeta$\,Tau obtained during science 
verification of the HR mode on October 7, 2008 were retrieved from the ESO Science 
Archive.  The three-telescope baseline configuration was UT1-UT2-UT4 for both stars.
The observing conditions were average to good for 48\,Lib (DIMM 
seeing $\sim$ 0.8--0.9\arcsec, coherence time $\sim$6.3\,ms), 
but quite poor during the $\zeta$\,Tau observations (DIMM seeing $\sim$1.2--1.5\arcsec, 
coherence time only $\sim$ 1.5\,ms). 

The central wavelength was set to 2.171958\,$\mu$m in the first order to match the 
Br\,$\gamma$ line.  The {\sc Finito} fringe tracker \citep{2008SPIE.7013E..33L} 
enabled us to record the whole spectral region with integration times of 3\,s and 
1\,s for 48\,Lib and $\zeta$\,Tau, respectively. HD\,143033 (diameter 1.36\,$\pm$ 
0.04\,mas) 
was observed as the calibrator after 48\,Lib and HD\,33833 (0.818$\pm$0.056\,mas) 
before $\zeta$\,Tau. The diameters of both calibrators were adopted from
the catalog by \citet{2005A&A...433.1155M}. The dates of the observations are 
given in Tab.\ ~\ref{amber_data}. u-v plots of the 48\,Lib {\sc PIONIER} and 
{\sc AMBER} observations are shown in Fig.\ \ref{uvplots}.

The reduction of the {\sc AMBER} data was performed using the yorick amdlib3.0 
package \citep{2007A&A...464...29T,2009A&A...502..705C}. The wavelength scale
-- affected by a problem with the positioning of the {\sc AMBER} grating --  
was corrected by fitting telluric lines using the Python procedure written by 
A. Merand.  
The same methods as in \citet{2009A&A...504..929S} were 
applied to derive absolute and differential quantities for visibilities, phases, 
and closure phases across the Br\,$\gamma$ line. Following the recommendation 
by \citet{2007A&A...464...29T} and \citet{2008eic..work..461M}, only the top 
50\% of the frames with the best $S/N$ were averaged for 
calibrators as well as  science targets to derive the transfer function and 
obtain absolutely calibrated spectra.  However, because of the setting inconsistency 
between the science target and the calibrator, only the differential visibilities 
and phases were derived for $\zeta$\,Tau. 

Differential visibilities, phases, and closure phase were averaged, and the 
values were normalized relative to the continuum in the intervals 
2.150--2.158\,$\mu$m and 2.178--2.190\,$\mu$m.  The differential method enables 
a more precise measurement of quantities with respect to the adjacent continuum.  
Compared to the absolute method, a significantly higher $S/N$ can be obtained 
because the atmosphere affects all spectral channels in nearly the same way.  
However, the absolute calibration of the angular size of the observed object 
is lost.

\begin{center}
   \begin{figure}[t]
   \includegraphics[angle=0,width=\columnwidth]{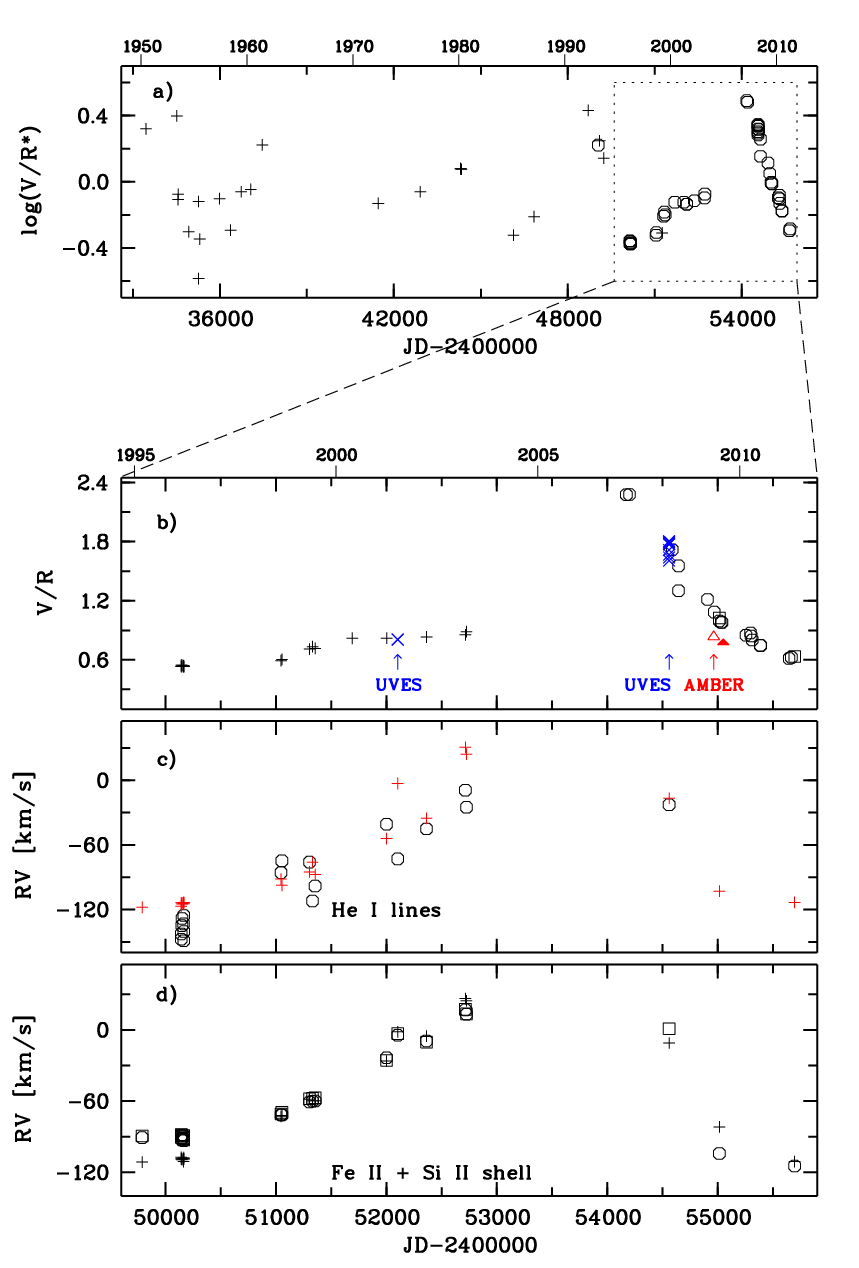}
   \caption{
a) History of 48\,Lib H$\alpha$ $V/R$ variations (defined as ratio of the peak 
intensities above the continuum) in the period 1950--2011. The circles mark 
original data from this paper and from \citet{2006A&A...459..137R}, + data
compiled from \citet{1953PDAO....9..363U}, \citet{1969A&A.....2..162F},  
\citet{1974ApJS...27..121G}, \citet{1978ApJS...38..205S}, \citet{1982A&AS...48...93A} 
\citet{1995A&A...300..163H} and \citet{2000A&AS..147..229B}. 
b) H$\alpha$ and Br$\gamma$ $V/R$ variations (defined as the ratio of the 
full peak intensities) during the period 1995--2010.  The arrows mark the dates 
of {\sc AMBER} and {\sc UVES} observations used for this study.  Data from different 
datasets are identified by the following symbols. H$\alpha$:  + \citet[][{\sc HEROS} 
spectrograph]{2006A&A...459..137R}, x VLT/{\sc UVES}, $\circ$ BeSS database, 
$\square$ LNA/{\sc Ecass}; Br$\gamma$: $\bigtriangleup$ VLTI/{\sc AMBER},  
$\blacktriangle$ Gemini/{\sc PHOENIX}. 
c) Radial velocities of selected \ion{He}{i} lines; mean of \ion{He}{i}\,4026 
and 5878 ($\circ$) and  \ion{He}{i}\,6678 (+). The RV error is 3--5\,\kms \ for
shell lines, but as large as 10--20\,\kms \ for \ion{He}{i} lines due to their 
peculiar line profiles.  
d) Radial velocities of \ion{Fe}{ii} ($\circ$ multiplet 96, $\square$ multiplet 117) 
and \ion{Si}{ii}, multiplet 2 (+).
}
\label{vtor}
\end{figure}
\end{center}

\section{Spectroscopic results}
\label{spectra}

\subsection{Long-term $V/R$ variations of emission lines}
\label{vr}

The $V/R$ ratio in our spectra was measured as the ratio of the emission peak heights 
(above the zero flux level) derived by fitting a Gaussian to the upper part of the 
peaks. 

Some $V/R$ studies \citep[e.g.][]{1997A&A...318..548O} report that
48\,Lib went through four $V/R$ cycles with lengths of about 13.5, 12, 7.5, and 
10 years between 1950 and 1994. The statement is supported mainly by Fig.\ ~1 in 
\citet{1995A&A...300..163H}, in which one can identify $V/R$ maxima  
close to 1950 and approximately in 1963, 1975, 1982 and 1992. However, the 
plot combines $V/R$ variations derived from H$\alpha$, H$\beta$ and H$\gamma$ lines
and for the 1975 maximum no H$\alpha$ values are plotted. As indicated also by the 
large differences between  H$\alpha$ and H$\beta$/H$\gamma$ values, combining $V/R$ 
values of different Balmer lines may result in a misleading picture both for  
observational and theoretical reasons: 
\begin{itemize}
   \item H$\beta$ and $\gamma$ are formed 
in the inner disk closer to the central star than H$\alpha$ and map different parts 
of the density wave spiral structure. Consequently, significant phase shifts between 
$V/R$ variations of different \ion{H}{i} lines can be expected - see 
\citet{2009A&A...504..929S} for $\zeta$\,Tau or \citet{1984PASJ...36..191K} for EW\,Lac.
   \item Intensities of the $V$ and $R$ peaks in H$\beta$ and $\gamma$ are much more 
affected by a strong stellar absorption. No correction using synthetic or template 
profiles was performed in the early observations.
\end{itemize}
For the above reasons a recompilation of the $V/R$ variations in H$\alpha$ $V/R$ only 
was performed and combined with the new 1995--2011 data. Results are shown in 
Fig.\ ~\ref{vtor}a. Because for some of the published datasets the authors used $V/R$ 
defined as $V/R^{\star} = (I(V) - I(cont)) / (I(R) - I(cont))$ and did not publish 
H$\alpha$ profiles, this definition had to be accepted also in Fig.\ ~\ref{vtor}a. 
Unlike in Fig.\ ~\ref{vtor}b, $\log (V/R^{\star})$ is plotted.

The plot reliably, although not precisely in time, shows maxima in 2005 and 1992. A 
weak maximum in 1980/1981 agrees with that in Hanuschik's plot. No H$\alpha$ 
observations were found in the literature for the period 1963--1976.  
Fig.\ ~\ref{vtor}a shows a few higher values around Hanuschik's maxima in 1950 
and 1963, but also a large scatter. All $V/R$ values before 1985 were measured from 
published H$\alpha$ profiles derived from photographic plates. We conclude that only 
two H$\alpha$ $V/R$ cycles are sufficiently documented in the star's history, giving 
cycle lengths of about 11 and 15 years. Only the last cycle, see Fig.\ ~\ref{vtor}b,c,d, 
will be analyzed in more detail in this paper.

Unfortunately, even for the last cycle, there are gaps  just when the 
quasi-periodic oscillations were probably absent (1995--1998) and re-started (at 
some time between 2004 and 2007).  Although detailed information about the 
H$\alpha$ behavior in these phases is, therefore, missing, 48\,Lib offers one of 
the not so frequent opportunities to study the recovery of $V/R$ activity in a Be 
star. 

Lower amplitudes seem to be connected with a shorter cycle length.  The observations 
obtained in 2000 show a first small increase in $V/R$ and thereafter almost constant 
$V/R\sim1.0$ for more than two years.  A strong increase to $V/R\gg1$ must have taken 
place at some time 
between 2004 and 2007.  Beginning with the first observation on JD\,54\,172 after 
this gap, $V/R$ continued to decrease and now reached the 1996 level of about 0.8. 
This places a lower limit on the length of the descending branch of the peak 
in the $V/R$ curve of about 4.2 years.  Taking into account the history of the
variations since 1970, see Fig.\ ~\ref{vtor}a, we can assume that the variability 
in Fig.\ \ref{vtor}b is cyclic.  From the speculative assumption of a 
symmetric curve, a $V/R$ cycle length of about 17 years could be derived.  This 
length would be significantly longer than the previous ones observed in the 
second half of the 20th century.  The present descending $V/R$ branch also suggests
-- consistently with the longer cycle --  
that the descent started from at least the same or even a higher value than the 
maximum in the previous activity cycle.

\begin{center}
   \begin{figure*}[t]
   \includegraphics[width=1\textwidth,angle=0,]{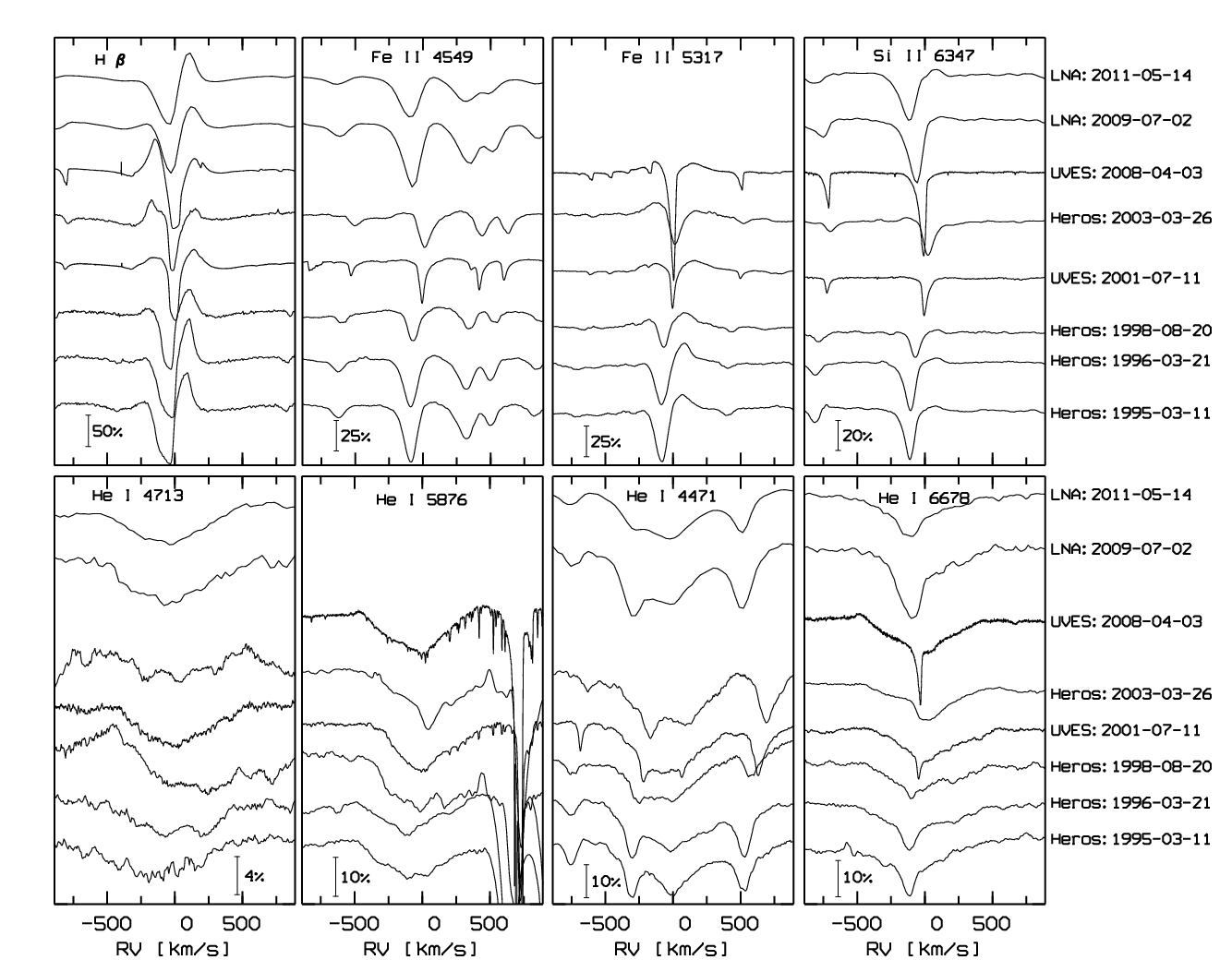}
   \caption{Long-term variations of H\,$\beta$ and selected \ion{Fe}{ii}, 
\ion{Si}{ii} and \ion{He}{i} line profiles obtained in 1995--2011.  Note the very 
different scale of the normalized flux, indicated at the bottom of each panel.  
All {\sc HEROS} and {\sc UVES} spectra were smoothed with a 5-pixel filter.  The 
spectrograph and date are marked in the right margin.
}
  \label{profiles}
  \end{figure*}
\end{center}

The upper left panel in Fig.\ ~\ref{profiles} shows that the H$\beta$ $V/R$ ratio 
follows a similar trend as the one of H$\alpha$. 

Apart from H$\alpha$ $V/R$ variations, Fig.\ ~\ref{vtor}b also includes Br\,$\gamma$ 
$V/R$ measurements made in spectra extracted from the {\sc AMBER} data and from 
the single {\sc PHOENIX} HR observation.  If the Br$\gamma$ $V/R$ curve exactly 
matches that of H$\alpha$, one could infer that Br$\gamma$ is ahead of H$\alpha$ 
by about 300--330 days.  For a mean cycle length of 15 years, this would correspond 
to a phase shift of 19--22\degr.

\subsection{Absorption line profiles}
\label{lp}

Shell absorption lines in Be stars are formed in the narrow sector of the 
circumstellar disk projected on the central star.  Since the involved range in 
azimuth is much larger along the stellar equator than in the more extended 
circumstellar disk, which moreover -- in the Keplerian case -- rotates much 
more slowly than the star, circumstellar absorption lines are mostly much narrower 
than photospheric lines.  In the simplest case of an axi-symmetric, Keplerian disk, 
the line profiles should be time-invariant.

The shell-star nature of 48\,Lib manifests itself through numerous strong and 
relatively sharp lines from singly-ionized metals in its spectrum.
Given the above basic model of Be-shell stars, their temporal variability is 
evidence of large deviations in the disk from a circular Keplerian radial 
velocity (RV) field with $v_\mathrm{rad}=0$\,\kms \ or a strong azimuthal asymmetry 
in the density distribution, or both.  Their 
profiles are an important indicator of the disk dynamics, which they sample in 
a complementary way to emission lines.   Furthermore, \citet{1996A&A...312L..17H} 
discovered extremely narrow multiple components of \ion{Fe}{ii} shell lines 
variable both in RV and depth on a timescale of less than one day.  The authors 
hypothesized that these features could be caused by co-orbiting clumps in the 
disk or by higher-order distortions of the global density wave causing the $V/R$ 
variations seen in the emission lines.

Fig.\ ~\ref{profiles} shows line profiles of H$\beta$ and some shell and 
\ion{He}{i} lines in selected epochs in 1995--2011.  The shell lines exhibit 
variations in depth and width, which are linked to $V/R$ (previous subsection) 
and RV variations (see next subsection) in H$\alpha$.  Both depths and widths of 
the shell lines reach a maximum close to the phase of minimal $V/R$ and RV. Owing 
to missing data, the observational correlation is not clear for $V/R>1$.  
The deeper lines may also show a blue-wing asymmetry as it would result from 
motions toward the observer.  It is worth noting that Fig.\ ~\ref{profiles} may 
be somewhat misleading because it is biased by the lower resolution of the OPD/LNA 
spectra.

The profiles of \ion{He}{i} lines are quite unusual.  The entire optical region 
covered by the {\sc UVES} and {\sc HEROS} spectra contains only two lines with 
profiles not 
obviously strongly affected by superimposed shell components, namely 
\ion{He}{i}\,4713 and \ion{He}{i}\,5786.  However, even the core of the latter 
is probably contaminated by circumstellar absorption.  In spite of the low S/N 
in \ion{He}{i}\,4713, Fig.\ ~\ref{profiles} illustrates that the line suffers from 
strong profile and asymmetry variations. 

\ion{He}{i}\,4471 and most other \ion{He}{i} lines show fairly complex profiles 
with one or more absorption components.  Most of them turn out to be superimposed 
metallic lines, e.g. \ion{Fe}{ii}\,4469.485 and 4472.086, or \ion{Fe}{ii}\,6677.305 
in the lines \ion{He}{i} 4471 and 6678, respectively.  However, because most 
\ion{He}{i} lines follow the same RV trend as the shell lines, one cannot exclude 
that their absorption cores are blends of metal and \ion{He}{i} lines.  Extended 
blue wings of \ion{He}{i}\,5876 and 6678 can be recognized during 
MJD\,52\,000--54\,000, while depressions in the red wing occurred in the 
interval MJD\,50\,000--51\,000 and around MJD\,55\,000. 

The observed line-profile variability strongly suggests that the above assumed 
distinction between photospheric and circumstellar lines is not nearly as simple 
as presented.  On the other hand, the profiles of lines formed in the disk are 
important line-of-sight probes of the disk dynamics, which they sample in a 
complementary way to emission lines.  The same diagnostics is not available for 
Be stars not viewed through the disk.  

The only datasets of this study with spectral resolution sufficient for the 
detection of very narrow (FWHM\,$\sim$\,10\,\kms) optical absorption components as 
reported 
by \citet{1996A&A...312L..17H} are those from {\sc UVES}.  However, no such narrow 
component could be detected in \ion{Fe}{ii} profiles.  A search in the {\sc HEROS} 
spectra, the very first of which was obtained only two days after those by 
\citeauthor{1996A&A...312L..17H} on March 10, 1995 (their Fig.\ ~1) also remained 
without positive results.  Because the spectral resolution of {\sc HEROS} at 
5300\,\AA\ is only $\sim$15\,\kms, one cannot decide whether the components were 
just not detected or really disappeared. A splitting of the absorption core or 
traces of extra absorption components can be seen in some \ion{Fe}{ii} lines 
in the {\sc HEROS} spectra obtained in 1996, but they cannot 
be reliably attributed to the same phenomenon.

\subsection{Radial velocities}
\label{rv}

Variable line profiles are quantitatively easiest characterized by radial-velocity 
(RV) measurements, especially if they have narrow cores.  Without higher-order 
perturbations, RVs  of photospheric and shell lines in Be 
stars should be the same.  However, \citet{2006A&A...459..137R} pointed out very 
high negative radial velocities of shell lines (-200\,\kms) in 48\,Lib although the 
star probably is not a binary (cf.\ Sect.\ref{discuss}).  Timescales of years are 
difficult to reconcile with any stellar processes.  Therefore, the variable radial 
velocities result from gas motions in the disk, and, as shown by 
\citet{1997A&A...318..548O}, 
the cycle lengths match periods of disk oscillations well.  But there is no 
straightforward interpretation of the RV of line cores in terms of particle 
trajectories or streaming motions because the line profiles are convolutions of 
velocity profiles with density and excitation profiles.  

To study the nature of the time-dependent RV field, radial velocities 
of \ion{Fe}{ii} and \ion{Si}{ii} lines were measured by fitting a Gaussian to 
the cores of their absorption profiles.  The results plotted in Fig.\ ~\ref{vtor}d 
show that H$\alpha$ $V/R$ and shell RV variations are well correlated.  The 
stagnation of the $V/R$ variations in 1996--2003 was accompanied by an increase 
in the RV of shell lines by about 100\,\kms.  In 2011, the values returned to the 
same negative velocity of about -100\,\kms \ as in the end of the 1990’s.  Separately 
plotted in Fig.\ ~\ref{vtor}d are mean velocities for \ion{Fe}{ii} multiplets 96 
(spectral lines at 4508, 4549 and 4583~\AA ) and 117 (lines 5197, 5234, 5275, 
and 5317\,\AA ) and \ion{Si}{ii} multiplet 2 (lines 6347, 6371\,\AA ).  In the LNA 
spectra, only lines of \ion{Fe}{ii} multiplet 96 and \ion{Si}{ii} could be 
measured. In spite of this inhomogeneity and different spectral resolutions, the 
values for different multiplets and for \ion{Fe}{ii} and \ion{Si}{ii} do not show 
significant differences.  Accordingly, the differentiation of the velocity field 
with excitation potentials may be small.  

The contamination of \ion{He}{i} profiles by metallic lines implies the 
potential risk that the \ion{He}{i} RVs are biased.
The risk was minimized by
\begin{itemize}
   \item Checking for the presence of metallic lines in the region of each \ion{He}{i}
line using a catalog of spectral lines \citep{1980wtpa.book.....R}. \ion{He}{i}\,6678, 
formed in the upper photosphere or even inner disk, as well as \ion{He}{i}\,4026 and  
\ion{He}{i}\,5876 were identified as the least contaminated ones according to both line
catalog and visual inspection of the lines.  The \ion {He}{i}\,4713 line was not 
included in the mean.  This line is very shallow so that RV measurements are strongly 
affected by noise and imperfectly corrected echelle ripple functions.
   \item Measuring \ion{He}{i} RVs in the line wings by fitting the profile between 
approximately one and two thirds of the core intensity by a Gaussian. The FWHM of the 
shell lines (variable, 50--80\,\kms \ as measured in HR {\sc UVES} spectra) is much lower 
than that of \ion{He}{i} lines (300--380\,\kms).  Provided that shell lines do not 
affect the line wings, they bring about a line profile distortion and a larger scatter
in RV, but they do not significantly affect the measured RVs. The fact that the RVs
of different \ion{He}{i} lines are consistent also excludes the possibility that they 
are mistaken for those of shell lines. The corresponding shifts of the entire broad 
\ion{He}{i} absorption profiles can also be recognized visually in Fig.\ ~\ref{profiles}. 
\end{itemize} 
Somewhat surprisingly, the RVs of selected \ion{He}{i} lines broadly agree 
with those of more classical shell lines.  Fig.\ ~\ref{vtor}c shows RVs of 
\ion{He}{i}\,6678 and the mean RV of \ion{He}{i}\,4026 and  \ion{He}{i}\,5876. 
 The scatter in the RVs of the 
He\,I lines is considerably higher than for shell lines.  Fig.\ ~\ref{profiles} also 
documents that line cores of all \ion{He}{i} lines are variable on a long 
timescale exhibiting wide ranges in shape and strength.  All these effects lead 
to errors in the RVs of \ion{He}{i} lines of a few tens of \kms.  

The errors are largest for the two lines that are supposedly mainly formed 
in the photosphere. \ion{He}{i}\,4026 shows two components, which are much broader 
than dynamical shell lines.  The blue component dominated the RV measurements in 
1995--2003 
and was included in the means plotted in Fig.\ ~\ref{vtor}.  Similar, but much more 
difficult to measure, components are present in \ion{He}{i}\,4009.

Apart from the large scatter, there are also considerable systematic differences 
between the RVs of individual lines.  Note that the values plotted in Fig.\ ~\ref{vtor} 
are not representative of the radial velocity of \ion{He}{i}.  They 
were merely included to illustrate some basic similarities to conventional shell 
lines.

\section{Imaging polarimetry}
\label{polar}

\citet{1999PASP..111..494M} performed a time-series analysis of his  broadband 
polarimetry covering more than a decade.  While the polarization angle, $\Theta$, 
was constant, the degree of the polarization, $p$, suggested quasi-cyclic 
variability, which McDavid fitted with a period equal to one-half of the cycle 
length of the H$\alpha$ $V/R$ cycle.  The peak-to-peak amplitude in $p$ was 
about 0.1\%.  A  correlation with the $V/R$ variability could be caused by the 
density variations 
associated with one-armed disk oscillations. The double-frequency variability of 
the polarization might arise because polarization cannot distinguish between the 
density enhancement being on the “left” or the “right” side of the disk.  
Unfortunately, the OPD/LNA polarimetry does not cover the $V/R$ stillstand, which 
might have provided an interesting test of the proposed, still unconfirmed 
interpretation.  

\begin{center}
   \begin{figure}[t]
   \includegraphics[width=\columnwidth]{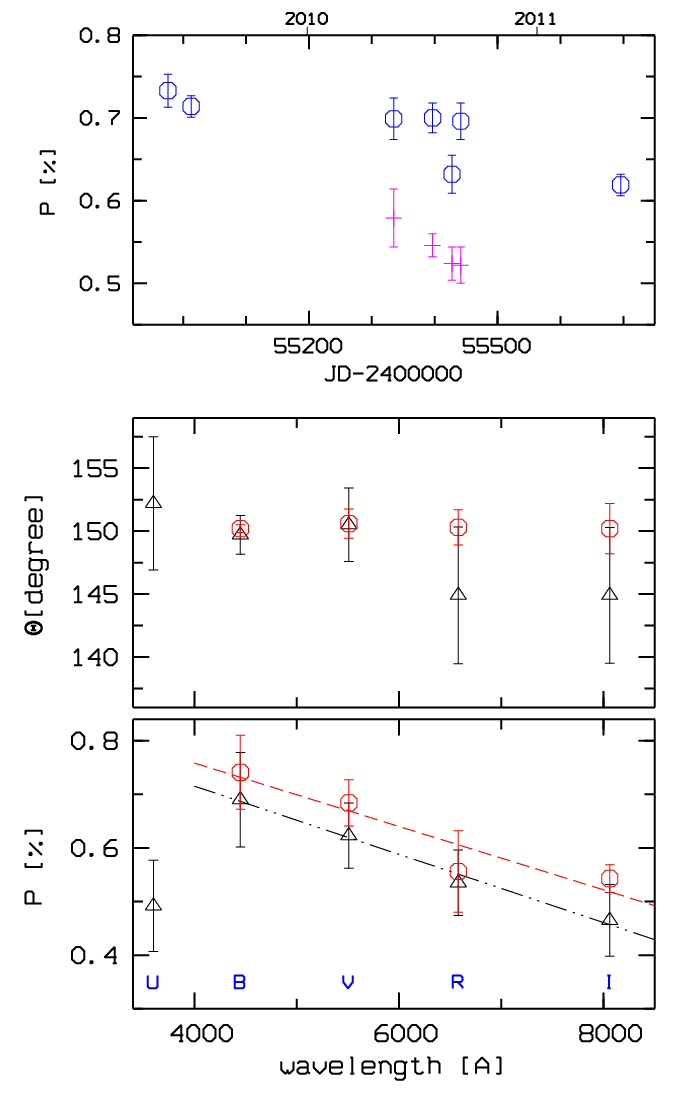}   
   \caption{48\,Lib polarization observations: Upper panel - decreasing $p$ trend in V 
    ($\circ$) and I (+) bands in the period 2009--2010. Lower panels -  mean 
    polarization and polarization angle vs. spectral band in the periods 1990--1998 
   (black $\triangle$) and  2009--2010 (red $\circ$). Straight lines in the bottom 
   panel show the regression  fits to the BVRI bands for both epochs.  Within the 
   errors the spectral slope does not change.   
   }
\label{com48pol}
\end{figure}
\end{center}

The upper panel of Fig.\ ~\ref{com48pol} shows a decreasing $V$ and $I$ band 
polarization trend in 2009--2010. A similar decrease is present also in the remaining 
spectral bands. The relative error of an individual observation is 2--4\%, while 
the drop of polarization in the given period is 10--15\%.  The lower two panels 
compare the mean values of $p$ and $\Theta$ for 48\,Lib in the period 1990--1998 
derived from the data of \citet{1994PASP..106..949M,1999PASP..111..494M} with the 
OPD/LNA observations obtained in 2009--2010.  $V$-band Pine Bluff Observatory 
data from 1989--1994 yield slightly lower values of $p$ than those derived from the 
McDavid dataset: $p(V)=(0.519\pm0.070$)\%  and  
$\Theta(V)=(150.3\pm1.0)^{\circ}$.  Because for an optically thin or moderately thick
disk, the polarization angle is always perpendicular to the disk plane 
\citep{1996ApJ...461..828W}, the value of $\Theta$ derived by McDavid is equivalent to 
the disk position angle (60.3$\pm$1.0)\degr.  
The errors show standard deviations, which are biased by real 
intrinsic variations as shown in the upper panel. Similarly, the 1990--1998 means 
are derived from seasonal averages and may be affected by variations on a timescale 
of years. 

Assuming overestimated standard deviations due to real trends, the difference between 
the mean polarization in the 90s and 2009--2010 can be considered as real and 
together with the decreasing temporal
trend  suggests that the polarization was increased at the time of the $V/R$ recovery
and in 2009--2010 it was returning to the level measured in the 90s. To within 
the errors, there is no change in the wavelength dependence. The maximum polarization 
appears in the B band.

\begin{center}
   \begin{figure}
   \includegraphics[angle=0, clip, width=0.9\columnwidth]{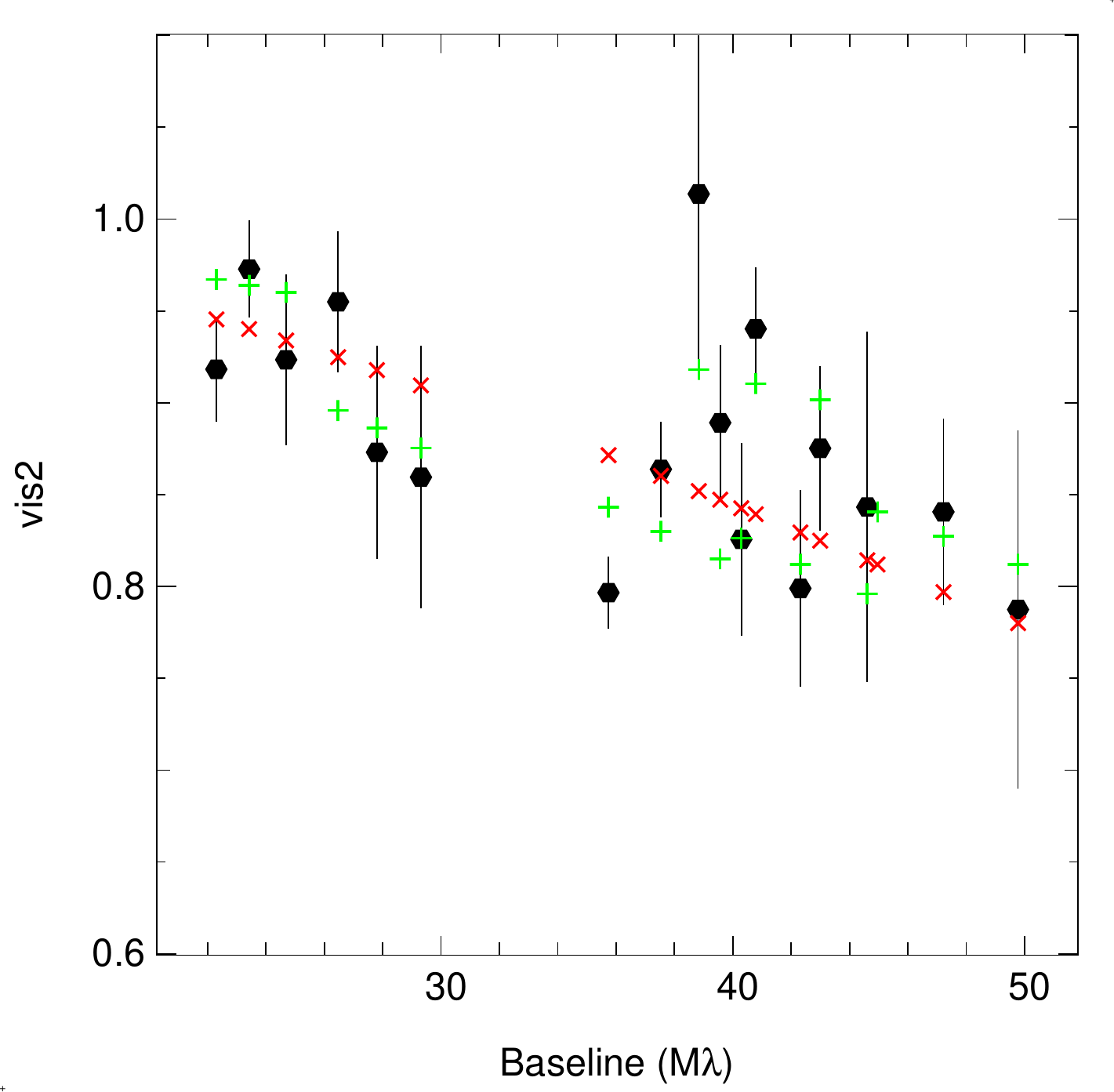}
   \caption{Absolute visibilities for 48\,Lib in the H band obtained with {\sc PIONIER} 
   and auxiliary telescopes at the D0-H0-G1-I1 VLTI baseline configuration. Black dots 
   mark the observed visibilities, crosses (x, red) the fit with a Gaussian disk, and plus 
   signs (+, green) the fit with an elongated disk.
   }
\label{pionierdata}
   \end{figure}
\end{center}

\section{Interferometry}
\label{intefer}

\subsection{Continuum}
\label{pionier}

The disk continuum emission around 48\,Lib is marginally resolved by the H broad-band 
{\sc PIONIER} interferometric observations. We studied its morphology via parameter 
modeling using the LITpro package \citep{2008SPIE.7013E..44T}. Results are summarized in 
Tab.\ \ref{pionierfit}. To facilitate the comparison with other published disk 
diameters, we first fitted the dataset with a uniform disk model, which yields
$\theta_\mathrm{UD} = (1.34\pm 0.05)\,$mas.  We then modeled more realistically the H band 
emission as a Gaussian profile disk and an unresolved component (the central star) that 
contributes two thirds of the total flux \citep{1994A&A...290..609D}. The derived 
FWHM of $\theta_\mathrm{GD} = (1.44\pm 0.05)\,$mas  is consistent with the results by 
\citet{2010ApJ...721..802P} (1.65 mas) obtained in the K-band continuum with the Keck 
interferometer.

Although an axisymmetric Gaussian disk plus an unresolved component reproduce the full 
{\sc PIONIER} dataset relatively well  ($\chi^\mathrm{2}_\mathrm{r}=2.07$), we explored the 
possibility to constrain the disk position angle. We obtained a  better fit ($\chi^2_r=1.33$) 
with an elongated Gaussian disk whose major axis has a FWHM of 
$\theta_\mathrm{GD}^\mathrm{major} = (1.72\pm 0.2)$\,mas 
and a position angle of $(50\pm 9)^{\circ}$ (north to east). The ratio between the FWHM of 
the major and the minor axis is 
$\theta_\mathrm{GD}^\mathrm{major}/\theta_\mathrm{GD}^\mathrm{minor}=1.66\pm0.3$.
Well within the errors, the polarimetric and the interferometric PA of the disk are 
consistent.

Finally, we explored the possibility of a fully resolved component that would be the 
signature of a dim shell with a spatial extent significantly larger than the disk 
itself. The resulting 
best-fit model has a negligible shell contribution of $0.01\pm0.02$ of the total flux 
and an increased $\chi^2_\mathrm{r}$ value compared to the model without shell. 
We conclude that a shell  
whose integrated contribution is more than a few percent in the H band is not probable 
around 48\,Lib.

\begin{center}
   \begin{figure}
   \includegraphics[angle=0, clip, width=0.9\columnwidth]{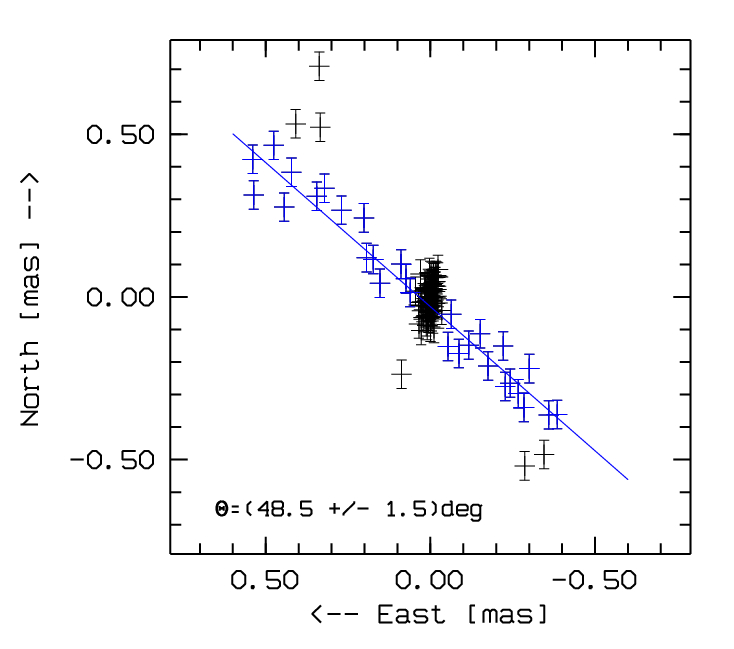}
   \caption{Relative photocenter shifts derived from the {\sc AMBER} differential
    phases across Br$\gamma$. After excluding the outliers at extreme offsets and points 
    corresponding to the continuum (-50$<$NE$<$50) the linear regression fit (blue line)
    to the photocenters across the line (plotted in blue) yields the disk position angle
    (48.5$\pm$1.5)\degr.  
    The NE shifted points correspond to the shorter-wavelength wing of the Br$\gamma$ 
    emission profile, the SW shifted points to its longer-wavelength wing. 
   }
\label{photocenter}
   \end{figure}
\end{center}

\begin{table}
\caption[]{Best fit for the parametric modeling of the H-band continuum emission from 
{\sc PIONIER} data. Position angles are defined north to east.}
\begin{tabular}{l|llll} 
  \hline \hline
  Model &  Parameter & Value & $\chi^2_r$ \\
  \hline
  Uniform Disk & $\theta_\mathrm{UD}$ & $1.34\pm 0.05\,$mas & 2.17
  \smallskip\\
  Star + Gaussian Disk & $\theta_\mathrm{GD}$ & $(1.44\pm 0.05)\,$mas & 2.07
  \vspace{1mm}\\
  Star + Elongated     & $\theta_\mathrm{GD}^\mathrm{major}$ & $(1.72\pm 0.20)\,$mas & 1.33 \\
  Gaussian Disk        & $\theta_\mathrm{GD}^\mathrm{major}/\theta_\mathrm{GD}^\mathrm{minor}$ & 
                       $(1.66\pm 0.3)\,$mas  & \\
                                    & $\mathrm{PA}^\mathrm{major}$ & $(50\pm 9)^{\circ}$ & 
  \vspace{1mm}\\
  Star + Elongated     & $\theta_\mathrm{GD}^\mathrm{major}$ & $(1.55\pm 0.20)\,$mas & 1.35 \\
  Gaussian Disk        & $\theta_\mathrm{GD}^\mathrm{major}/\theta_\mathrm{GD}^\mathrm{minor}$ & $1.93\pm 0.60$ & \\
  + dim shell          & $\mathrm{PA}^\mathrm{major}$ & $(53\pm 10)^{\circ}$ & \\
                       & $\mathrm{excess/total flux}$ & $0.01\pm 0.02$ &\\
  \hline\noalign{\smallskip}
\end{tabular}
\label{pionierfit}
\end{table}

\subsection{Br$\gamma$ line}
\label{amber}

The spectro-interferometric observations of 48\,Lib  were carried out with {\sc AMBER}
during the descending  branch of the $V/R$ activity (see arrows in Fig.\ ~\ref{vtor}). 
Because Fig.\ ~\ref{vtor} does not cover the 48\,Lib $V/R$ maximum and the cycle 
length is much more variable than in $\zeta$\,Tau \citep{2009A&A...504..929S}, the phase 
estimate is little 
accurate. One can, however, estimate that the {\sc AMBER} observations took place 
close to phase 90\degr since V$\simeq$R. This is also the approximate phase 
of the {\sc AMBER} observations of $\zeta$\,Tau, which justifies the inclusion of 
the latter data for comparison.  

\begin{center}
   \begin{figure*}
   \includegraphics[angle=0, clip, width=0.9\textwidth]{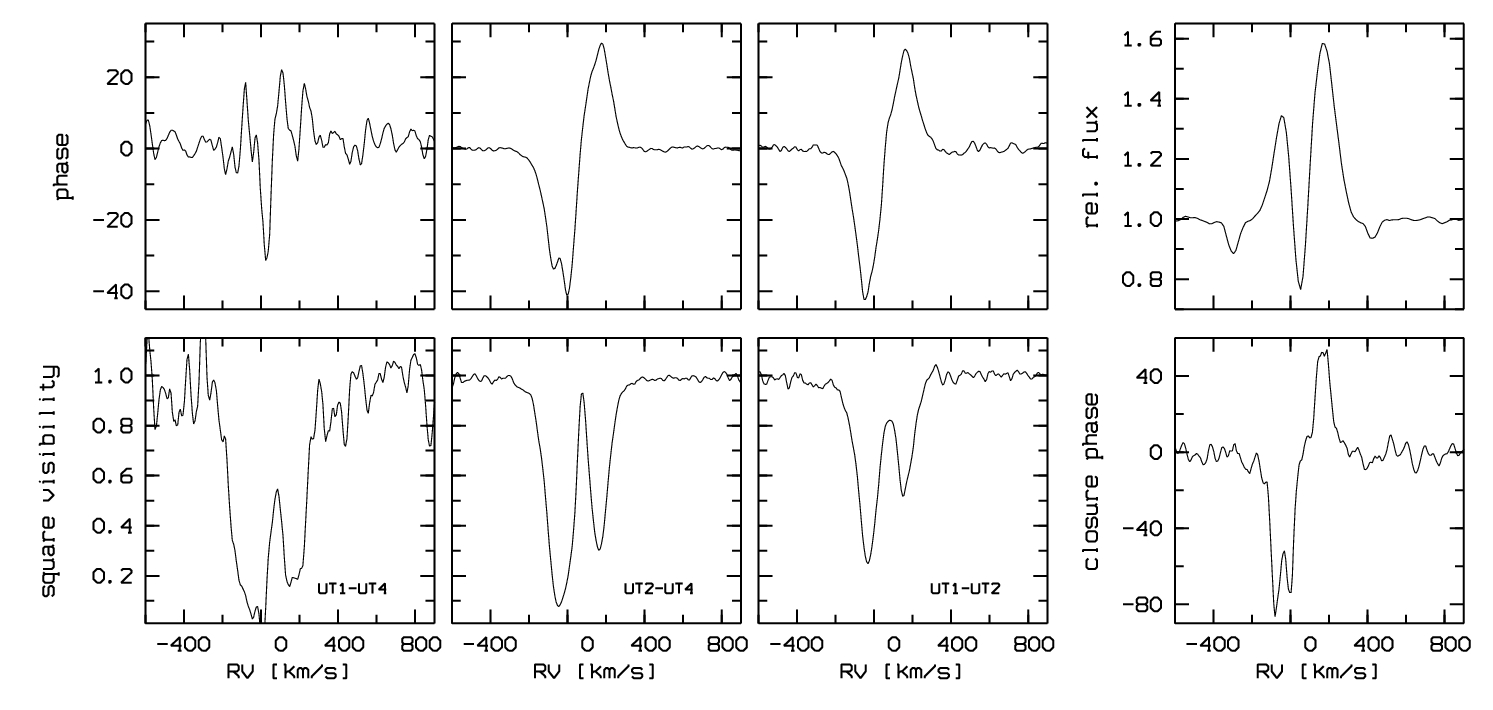}  
   \caption{Differential visibilities, differential phases, and closure 
   phase for 48\,Lib.  Lengths and position angles of individual baselines are: 
   UT1-UT2 56.6m@26.2\degr, UT2-UT4 89.4m@81.3\degr, and 
   UT1-UT4 130.2m@60.4\degr.
           }
\label{vis48}
   \end{figure*}
\end{center}

\begin{center}
   \begin{figure*}
   \includegraphics[angle=0, clip, width=0.9\textwidth]{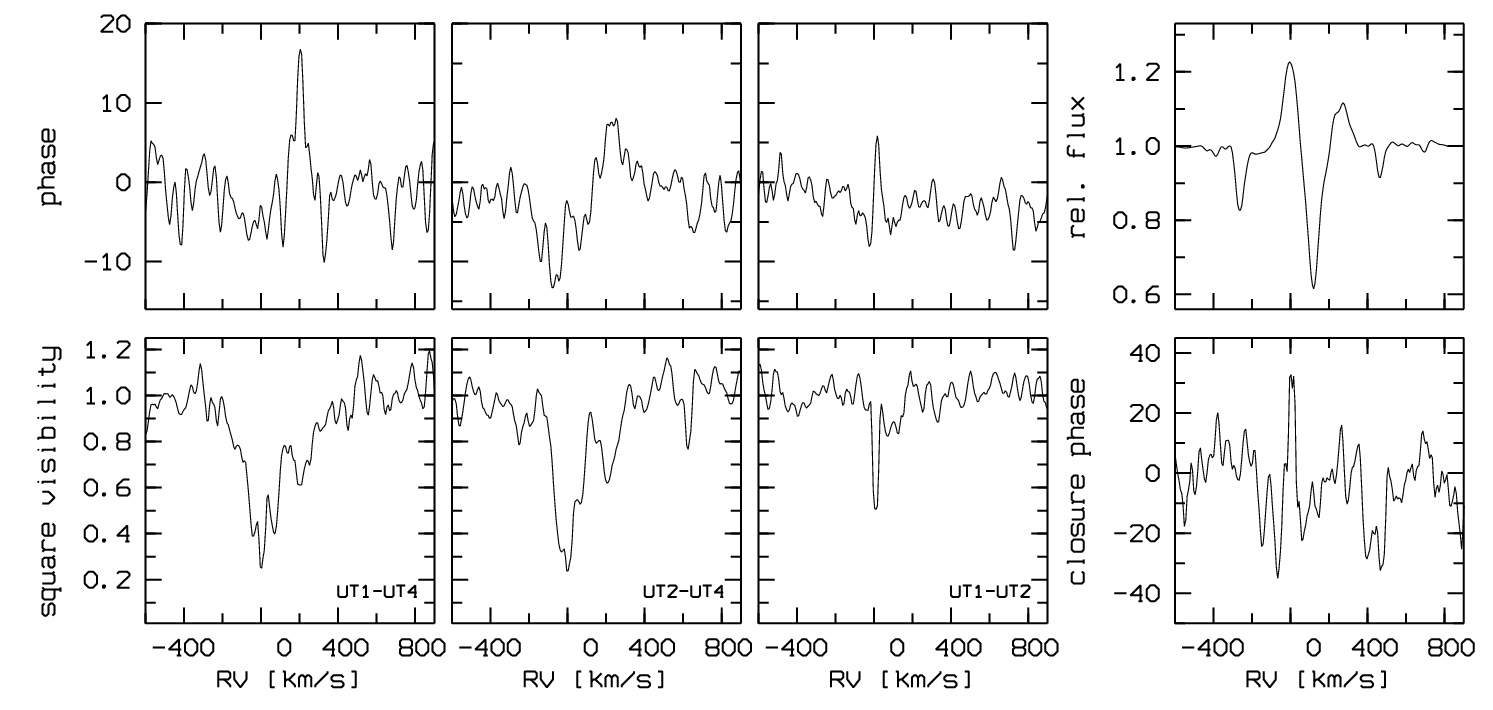}
   \caption{Differential visibilities, differential phases, and closure phase for 
   $\zeta$\,Tau. The baselines were the same as for 48\,Lib (cf.\ 
   Fig.\ ~\ref{vis48}).
           }
\label{viszt}
    \end{figure*}
\end{center}

The high-quality observations of 48\,Lib permit good-fidelity differential 
visibility and phase profiles to be derived (Fig.\ ~\ref{vis48}).  The disk was 
well-resolved at all three baselines, both in continuum and line emission.  
At the UT1-UT4 baseline, the disk was almost over-resolved in Br$\gamma$.
Although the data were absolutely calibrated, the analysis of continuum visibilities 
yields results inconsistent with {\sc PIONIER} continuum visibilities and with polarimetric 
observations. Consequently, only differential visibilities and phase profiles over 
Br$\gamma$ are discussed below.

For disk emission lines, simple uniform or Gaussian disk models predict a 
single-peak central reversal (akin to a spectral-line flux profile with 
superimposed circumstellar emission line) in the visibility and a smooth S-shaped 
phase profile.  Indeed, Fig.\ ~\ref{vis48}, baseline UT1-UT2, exhibits such an 
S-shape phase profile.  However, overall the 48\,Lib HR {\sc AMBER} profiles are 
significantly different.  All Br$\gamma$ visibility profiles show a double-component 
structure with a central peak and an extra peak, superimposed to the S-shape phase 
profile conventionally seen at lower spectral resolution. 
 
The height of the central peak in the 48\,Lib visibility function reaches the level 
of the continuum at the intermediate baseline and about half of it at the longest 
and shortest one.  
The central visibility peak seen at the UT1-UT4 baseline is accompanied by a 
multi-peak phase profile.  At the UT2-UT4 baseline, only a very weak phase reversal 
emerged in the negative lobe of the S, and the UT1-UT2 baseline shows a pure 
S-shaped profile (see Fig.\ ~\ref{vis48}).  

The low-noise 48\,Lib data also reveal a rich structure in the closure phase 
profile.  The high amplitude of the closure phase indicates a strong asymmetry.  
However, the splitting into both positive and negative peaks is well above the 
noise level. 

In spite of the higher noise, a similar double-component structure can be seen in 
the $\zeta$\,Tau visibility and phase profiles (Fig.\ ~\ref{viszt}).  The central 
peak in $V^2$ almost reaches the continuum level at the UT2-UT4 baseline.  Unlike 
in 48\,Lib, the visibility shows a gradual increase in the longer-wavelength wing 
and an asymmetry between the two visibility components.  For $\zeta$\,Tau, both 
interferometry and polarimetry give a disk major axis position angle of 32\degr 
 \citep{2009A&A...504..929S}. This is almost parallel to the UT1-UT4 baseline and 
the difference in position angle to the other two baselines is only about 
30\degr.  Although none of the baselines was close to perpendicular in the 
October 2008 {\sc AMBER} observations, Fig.\ ~\ref{viszt} still indicates some 
features over and above the basic S-shape.  But their significance should be 
verified by better quality data with different baseline angles.  

Not only are the HR {\sc AMBER} observations of 48\,Lib and $\zeta$\,Tau quite 
similar, they are also fully consistent with comparable observations of five 
other classical Be stars discussed by \citet{NMworkshop}. 

The differential phase signal can be converted into the 2D astrometric shift 
$\vec{p}(\lambda)$, which defines shifts of the photocenters in individual wavelength 
bins across the spectral line and the orientation of the disk major axis projected 
on the sky. Here we consider the applicability of the photocenter approach to 48\,Lib 
and other Be stars. The photocenter can be interpreted using the Zernike-van Cittern 
theorem and moments of the flux distribution \citep[][see their Eq. 4 for the 
definition of the nth moment]{2003A&A..400..795L}. In this formalism, the phase of 
the interferometric signal is proportional to the first flux moment, defined as the 
photocenter vector, and a third-order term, related to the object symmetry. The symmetry 
is described by the closure phase (Lachaume's Eqs. 8b and 10). If the closure phase 
is close to zero, the conversion from the phase to the astrometric shift can be  
made by inverting the formula 
\begin{equation} 
\phi(\lambda) = -2\pi \cdot \vec{p}(\lambda) \cdot \frac{\vec{b}}{\lambda} \hspace{5mm} ,
\end{equation}  
\vspace{0.07cm} 
\noindent where $\vec{b}$ is the interferometric baseline vector projected onto 
the sky. Because the phase does not depend on the second moment in the first
approximation a sufficient condition for the conversion is a negligible closure 
phase, not necessarily a marginal visibility, as required for the mean diameter or 
symmetry expressions.

A density wave in the Be star disk obviously leads to asymmetries and 
moderate or even large closure phase, see Fig.\ ~\ref{vis48} for 48\,Lib. Consequently,
Eq. 1 can be used only as a rough estimate of the disk orientation. 

Fig.\ ~\ref{photocenter} shows the relative photocenter offsets across the Br$\gamma$ 
line with respect to the continuum photocenter.  Although it over-constrains the 
problem, the astrometric shifts were computed for all three baselines and their 
consistency was checked.  The 2-D photocenter was reconstructed from the two shortest 
baselines (UT1-2, UT2-4) with the lowest noise and sufficiently different PA, see 
Fig.\ ~\ref{uvplots} . The UT1-4 baseline is inconsistent.  The relatively good 
agreement of the disk position angles derived from the {\sc PIONIER} fit and the 
{\sc AMBER} photocenter shift indicates that the shorter baselines in {\sc AMBER} 
observations were still close to the astrometric regime.
The approaching part of the rotating disk (the blue wing in the Br$\gamma$ line)
is oriented in the NE direction, the receding part to the SW.

\section{Discussion}
\label{discuss}

Observations of 48\,Lib between 1995 and 2011 document a recovery of the $V/R$ 
activity and suggest a correlation between the underlying density wave and RVs and 
profiles of metallic shell and \ion{He}{i} lines.  During a long plateau in the 
$V/R$ ratio through $\sim$2003 the RV still continued to increase so that an 
absence of detectable $V/R$ variations does not exclude ongoing global disk 
oscillations.  

The one-armed disk oscillation model has so far not been used to simulate a case 
like the one observed in 48\,Lib with substantial asymmetries of shell lines 
apparently caused by very large shifts of their cores.  Such high RV amplitudes 
of shell lines were also observed in other Be-shell stars, e.g., $\zeta$\,Tau 
\citep{2006A&A...459..137R}. Nevertheless, 
in the model of \citet{2009A&A...504..915C} for the Be star $\zeta$\,Tau, the 
particles orbit the star in very eccentric elliptical orbits. In that particular 
model, the radial velocities are as high as $100\,\rm km\,s^{-1}$.
Elliptical orbits could also explain the 
strongly variable width of metallic shell lines.  Obviously, such large intrinsic 
variations are much more readily observed in disks with an edge-on orientation 
towards the observer, i.e.\ in Be-shell stars.  

48\,Lib is also of special interest because many 
of its \ion{He}{i} lines also exhibit RVs in line wings, which broadly track those of 
shell lines.  Because circumstellar components of \ion{He}{i} lines must form very 
close to the central star, they are a very useful diagnostic of the inner structure 
of the disk.  Observations of Be stars \citep{1998A&A...333..125R} 
and modeling of their light curves \citep[][]{2011arXiv1112.0053C} 
provide increasing evidence for a time-dependent mass decretion rate and an 
episodic refilling of the disk.  After the decretion is stopped or significantly 
reduced, the time scale to stabilize the inner disk in the resulting 
low-density state is only on the order of weeks to months.  The observed strengths 
and velocities of the \ion{He}{i} line cores suggest that a high density is 
maintained in the inner disk through a high decretion rate.  
The large outer disk radius estimated from interferometry would be in line with 
this interpretation. 

Extended blue wings of \ion{He}{i}\,5876 and 6678 as observed during 
MJD\,52\,000--54\,000 
could also be a mass-loss indicator.  However, the observed dependency on $V/R$ 
phase, i.e., azimuth angle of the disk, makes this less likely.  An alternative
interpretation is offered from our preliminary modeling of the disk dynamics.  In 
some $V/R$ phases, the line-of-sight radial velocity shows 
an extra minimum in the very inner disk. Such a  
velocity field  could then be responsible for the extended blue wings 
and even double shell profiles, which were observed close to $V=R$ in $\zeta$\,Tau 
by \citet{2009A&A...504..929S} [their Fig.\ ~3 and related discussion]. 

In Be double stars, orbital and $V/R$ periods are in some cases known to be 
synchronized \citep{2007ASPC..361..274S}.  Their orbital periods are on the order 
of tens to hundreds of days, e.g. 4\,Her: 46.18\,d \citep{1997A&A...328..551K} or 
$\epsilon$\,Cap: 128.5\,d \citep{2006A&A...459..137R}.  For a putative binarity of 
48\,Lib, Fig.\ ~\ref{vtor} would be consistent with a 
period of 15--25 years.  The shape of the RV curve would imply only minor 
eccentricity.  Therefore, a binary would stay far away from the disk at all times 
and the very extreme disk variability could hardly result from the presence of a 
companion star, making a binary model very unattractive for 48\,Lib. 

Three independent methods -- polarimetric angle, LITpro modeling of continuum visibilities
and astrometric analysis of the photocenter shifts  -- yield the position angle of the 
major disk axis in the interval 48\degr to 60\degr.
It is worth mentioning that each of the methods may suffer from some systematic errors.
The polarimetric angle of 48\,Lib depends strongly on the correction for the interstellar 
polarization. The elongated axially symmetric disk assumed in the LITpro solution can be 
considered to be only the first-order approximation.  
The astrometric analysis was applied under conditions that do not guarantee the 
convertability 
from the interferometric phase of the used {\sc AMBER} observations and must be 
considered only as a rough estimate. The three methods provide information 
on parts of the disk at different distances from the central star. While the 
polarization comes from the inner disk region, which is also close to the region 
of formation of the H-band continuum, the Br$\gamma$ line is formed much farther 
out in the central part of the disk.  Considering all the above facts, the data 
do not exclude the possibility that the position angles of the inner disk and the 
H-band continuum and Br$\gamma$-emitting regions are nearly identical.

The accuracy of the {\sc AMBER} absolute visibilities reaches about 5\%
under optimal conditions  \citep{2009A&A...502..705C} and can be estimated between
6 and 10\% in our 48\,Lib observations. The inconsistency of the {\sc AMBER} absolutely
calibrated continuum visibilities with those from {\sc PIONIER}, polarimetric 
observations as well as simple LITpro models indicate other problems with the 
absolute calibration of the {\sc AMBER} HR continuum visibilities. Possible,
but so far little investigated sources of systematic errors are the fringe tracker 
{\sc FINITO} and integration times as long as several seconds. Note that the test 
by  \citet{2009A&A...502..705C} was made only for {\sc AMBER} medium resolution and
integration times between 0.02 and 0.16\,s.

A reliable calibration of the continuum visibilities could be made for 
the {\sc PIONIER} H continuum visibilities. Their fitting by the relatively simple 
LITpro disk models confirms the disk diameter of 1.7\,mas determined by the Keck H
broad-band interferometry and the position angle of 60\degr derived from our 
polarimetric observations.  This diameter at the Hipparcos parallax of 6.97\,mas 
corresponds to 0.25\,AU. Assuming a B3V main-sequence star radius of 3.56\,\Rsun \
\citep{1988BAICz..39..329H}, the H continuum disk diameter is about 15\,\Rstar. 
According to Fig.\ ~1 in \citet{2011IAUS..272..325C}, the K-band continuum is 
formed typically at about 1--3\,\Rstar, but Br$\gamma$ only about three times 
farther from the central star. Consequently, we can assume  that the diameter of 
the  Br$\gamma$ emitting region might be at least 30--50\,\Rstar. 

The LITpro estimate confirms that the 48\,Lib cicumstellar disk 
has one of the largest angular diameters among Be stars.  A consistent 
modeling of the 48\,Lib disk using more comprehensive interferometry and spectral 
energy distributions is under consideration.  It would also test the Keplerian 
dynamics assumed by the disk oscillation model, for which the extreme radial 
velocity amplitudes of shell lines may still be a challenge.  

Only thanks to the {\sc AMBER} high-spectral resolution mode could the 
double-component structure of visibility and phase profiles be discovered. In 
48\,Lib, the Br$\gamma$ line width is about 45\,\AA \ so that at $R=12\,000$ about 
25 wavelength bins sample the central visibility maximum very well and can reveal 
even more subtle features in the visibility profile.  At {\sc AMBER}'s medium 
spectral resolution ($R\sim1500$) just three wavelength bins cannot establish the 
same structure.  In spite of the very high phase stability, no double peak 
structure or traces of a central peak can be seen in visibility profiles of 
\citet[][their  Fig.\ ~2]{2010ApJ...721..802P}, obtained about a year before the 
{\sc AMBER} data and at $R=2000$.  Spectral resolution is also of key importance 
for interferometric studies of the disk dynamics.

The occurrence of similar visibility and phase profiles in the more $V/R$-stable 
disk of $\zeta$\,Tau as well as other stars \citep{NMworkshop} shows that they do 
not result from a dynamical peculiarity related to the $V/R$ recovery in 48\,Lib.  
Instead, these structures are a general property independent of the phase of disk 
evolution. 

Comparison with the 48\,Lib Br$\gamma$ emission profile offers a simple qualitative 
explanation of the W-shape visibility profiles. The Br$\gamma$ visibility is much 
reduced in the line center, where the line emission drops and the visibility is 
given mainly by the continuum visibility. \citet{2011arXiv1109.3447K} analyzed 
HR interferometric observations of the Be stars $\zeta$\,Tau and $\beta$\,CMi 
and proposed that phase reversals correspond to the transition from the first 
to the second visibility lobe. This over-resolution effect can explain the phase 
reversals in the case of large and close disks like that of 48\,Lib. This  
interpretation is consistent 
also with the fact that the phase reversal at the UT2-UT4 baseline and the second 
closure phase minimum appear at the position of the first and deeper visibility 
minimum. 
However, the lobe transition may not provide a general explanation of the phase 
reversal, particularly for less resolved and distant disks. \citet{NMworkshop} 
speculated that phase reversals and multi-peak closure phase profiles can be 
caused by secondary effects in the disk and may hint at additional morphological 
disk features. 
However, until a realistic modeling of the disks including the density waves is 
performed, the general explanation of the HR phase and closure phase profiles 
remains open. 


\section{Conclusions}
\label{concl} 

The compiled spectroscopic data of 48\,Lib confirm that its circumstellar disk is 
evolving on a timescale of years to decades.  The correlation between the 
H$\alpha$ $V/R$ ratio and RVs of the cores of metallic shell and even \ion{He}{i} 
lines implies a dense inner disk fairly close to the photosphere or even touching 
it.  An elongated disk model fit of the broad-band interferometric data indicates an 
outer disk radius as large as 15\,\Rstar in the H continuum. This leads to an 
estimate of the Br\,$\gamma$ radius  of several tens of stellar radii. 
Together with a relatively large parallax of 
6.97\,mas, these results make 48\,Lib a prime candidate for 
spectro-interferometric studies of the dynamical structure of Be-star disks.   
A more exact model would also shed light on the question 
whether the high-velocity cores of shell and \ion{He}{i} lines find their 
explanation within the density wave model alone or if an episodic gas outflow to, 
or infall from, the circumstellar disk must be included.  

The resumption of the $V/R$ oscillations in the first decade of this century was 
accompanied by a deepening of the disk shell profiles and an increased polarization 
degree.  

For more than 20 years, the polarization angle has been constant to within the errors 
and remained unaffected by the disk oscillation, which, therefore, must be purely 
planar without major 3-D precession or warping. 

At high spectral resolution, interferometric visibility and phase show significant 
deviations from the simple S-shape profile conventionally adopted for the 
interpretation of interferometric data of Be stars at low and medium spectral 
resolution. Only a realistic modeling  (Faes et al., in preparation) will be able
to answer the question if double-component phase or closure phase
profiles are always only a consequence of interferometric effects during visibility
lobe transitions or if they (also) reflect secondary effects in the disk.

\begin{acknowledgements}
This work has made use of the BeSS database, operated at LESIA, Observatoire de 
Meudon, France: http://basebe.obspm.fr. Particularly, 
we appreciate the effort of amateur astronomers M.\ Bonnement, R.\ Buecke, C.\ Buil, 
V.\ Desnoux, T.\ Garrel, J.\ Guarro-Fl\'{o}, F.\ Houpert,  B.\ Mauclaire,  E.\ Pollmann, 
J.\ Ribeiro,  J.\-N.\ Terry and S.\ Ubaud. 

The {\sc HEROS}@Ond\v{r}ejov monitoring
was part of a joint project supported by the German Bundesministerium f\"{u}r
Bildung und Forschung and the Ministery of Education of the Czech Republic
(TSE–001–009, ME–531) as well as the Deutsche Forschungsgemeinschaft and
the Academy of Sciences of the Czech Republic (436 TSE 113/18 and 41). S\v{S}
also appreciates the support of the Academy of Sciences and Grant Agency of
the Academy of Sciences of the Czech Republic (AA 3003403, K2043105) in the 
period of the {\sc HEROS} observations. Special thanks go to O.\ Stahl and M.\ Maintz
for reducing the {\sc HEROS} spectra, J.\ Wisniewski for providing us with the value 
of the 
interstellar polarization toward 48\,Lib and A. Merand, who gave his Python code for 
the wavelength correction to our disposal prior to its inclusion in the amdlib3 library. 

This research has made use of the  \texttt{AMBER data reduction package} of the 
Jean-Marie Mariotti Center\footnote{Available at http://www.jmmc.fr/amberdrs},  
\texttt{LITpro} service co-developped by CRAL, LAOG and FIZEAU \footnote{LITpro 
software available at http://www.jmmc.fr/litpro} and the SIMBAD database, operated 
at CDS, Strasbourg, France.  ACC acknowledges support from CNPq 
(grant 308985/2009-5) and Fapesp (grant 2010/19029-0)

This study is
partly based on observations obtained at the Gemini Observatory, which is operated 
by the Association of Universities for Research in Astronomy, Inc. (AURA), under a 
cooperative agreement with the NSF on behalf of the Gemini partnership: the 
National Science Foundation (United States), the Science and Technology Facilities 
Council (United Kingdom), the National Research Council (Canada), CONICYT (Chile), 
the Australian Research Council (Australia), Minist\'{e}rio da Ci\^{e}ncia e 
Tecnologia (Brazil) and Ministerio de Ciencia, Tecnolog\'{i}a e Innovaci\'{o}n 
Productiva (Argentina).  The data presented in this paper originate from Gemini 
programme GS-2009B-Q-93, observed in queue mode.

We thank the unknown referee for the constructive comments.

\end{acknowledgements}
 
\bibliographystyle{aa} 
\bibliography{48Lib.bib} 
\end{document}